# Characterization of ion/electron beam induced deposition of electrical contacts at the sub-µm scale

D. Brunel[1], D. Troadec[1], D. Hourlier[1], D. Deresmes[1], M Zdrojek[2] and T. Mélin[1,a]

[1]*Institut d'Electronique, de Microélectronique et de Nanotechnologie, CNRS UMR 8520, Avenue Poincaré, F-59652 Villeneuve d'Ascq, France.*

[2]*Faculty of Physics, Warsaw University of Technology, Koszykowa 75, 00-662 Warsaw, Poland.*

[a] Electronic mail: thierry.melin@isen.iemn.univ-lille1.fr

## Abstract

We investigate the fabrication of electrical contacts using ion (IBID) - and electron-beam induced deposition (EBID) of platinum at the sub-µm scale. Halos associated with the metal surface decoration in the 0.05-2 µm range are characterized electrically using transport measurements, conducting atomic force microscopy and Kelvin force microscopy. In contrast with IBID, EBID electrodes exhibit weakly conductive halos at the sub-µm scale, and can thus be used to achieve resist-free electrical contacts for transport measurements at the sub-µm scale. To show this, four-point transport measurements using µm-spaced EBID contacts are provided in the case of a multiwalled carbon nanotube.

Low-dimensional nanostructures such as carbon nanotubes, semiconductor nanowires, or graphene nanoribbons [1,2] are promising materials for nanoelectronic applications. Their dimensions allow the improvement of electronic properties of devices [3-5], and, for example, the fabrication of efficient electronic or biological sensors [6,7]. The most common technique to connect nanostructures is electron beam lithography (EBL) [8], which enables routine fabrication of electrical contacts with sub-100 nm gap spacing. However, depending on experimental situations, EBL may suffer from two disadvantages: (*i*) it is a process which requires multiple technological steps; (*ii*) the use of electron-beam resist and lift-off chemicals is not compatible with some material classes such as *e.g.* conjugated polymer nanowires [9,10]. This is why direct metal deposition techniques excluding resist patterning are being developed, such as *e.g.* nanostencil [11] - derived from atomic force microscopy -, and ion-beam induced metal deposition [12] as a technique based on focused ion beam (FIB) microscopy.

Here we focus the use of ion-beam induced deposition (IBID) or electron-beam induced deposition (EBID) of metal contacts. The principle of IBID and EBID is based on the local deposition (with resolution better than 100 nm) of metallo-organic compounds (usually containing Pt or W) under a focused ion- or electron beam, respectively. In spite of early promises as a direct-writing method to connect nanostructures [13], it has soon appeared that the use of IBID and EBID are limited by several factors, such as: a FIB-induced amorphization of the nanostructure below the contacts in IBID [14,15]; a change of the nanostructure transport properties upon a FIB beam exposure up to a ~10 µm distance from the contacts [16]; a larger resistivity of EBID contacts as compared to IBID, even after contact annealing [13,17]; and the formation of halos and leakage pathways around IBID and EBID contacts [13], associated with a metal decoration of the sample surface, up a ~10 µm scale in EBID [18]. These features have so far restricted the use of IBID and EBID to the fabrication of electrical contacts with minimum 5-10 µm gap spacing [13,16].

In this Letter, we investigate the fabrication of electrical contacts using IBID and EBID techniques at the sub-µm scale. Halos associated with the metal surface decoration in the 0.05-2 µm range are characterized electrically using transport measurements, conducting atomic force microscopy and Kelvin force microscopy. Our results confirm that IBID is not suitable for electrical contacts due to the large leakage, but demonstrate that EBID can nonetheless be used at the sub-µm scale, in conjunction with four-point transport measurements to account for the high-resistivity of the deposited contact patterns. Transport measurements using µm-spaced EBID contacts are performed on multiwalled carbon nanotube.

Experiments have been conducted with a FEI Strata dual-beam DB235 FIB system, enabling sample imaging in scanning-electron microscopy (SEM) mode, as well as IBID and EBID platinum deposition. For this a trimethylcyclopentadienyl-platinum $(CH_3)_3CH_3C_5H_4Pt$ mettalo-organic precursor gas has been used. To characterize IBID and EBID deposited electrodes (conductivity and leakage paths), we first used test samples fabricated by EBL, consisting of pre-patterned 25 nm thick gold electrodes with 5 µm gap spacing laying on a 200 nm thick $SiO_2$ layer thermally grown from a doped silicon substrate. Experiments on contacted carbon nanotubes have been performed using predefined EBL electrodes separated by ~10 µm on similar substrates. Individual nanotubes (here commercial multiwalled nanotubes from Nanocyl, Belgium) have been first deposited from a dispersion in dichloromethane, and then directly contacted either by IBID or EBID. Scanning-probe experiments have been conducted on a Dimension / Nanoscope IV atomic force microscope (AFM) from Veeco Instruments, operated either in conducting-AFM (c-AFM) or Kelvin Force Microscopy (KFM) modes. We used EFM PPP (Nanosensors) tip with low spring constants (a few N/m) enabling tapping- and contact-mode topography scans as well as KFM and c-AFM data acquisition. Four-point transport measurements have been carried out using an Agilent 4155 semiconductor parameter analyzer.

The electrical properties of the deposited Pt-based materials are first characterized by measuring the resistance through IBID (Fig. 1a) and EBID (Fig. 1b) bars deposited between predefined EBL Au contacts with 5 µm gap spacing. Bars with nominal size 8×0.5 µm² (IBID) and 7×0.5 µm² (EBID) have been deposited, with the following FIB IBID (and EBID respectively) operating conditions: 10 kV (5 kV) acceleration voltage, 10 pA (0.4 nA) $Ga^+$ (electron) beam current, 1 µs (0.5 µs) dwell time, and 18 s (600 s) total exposure time. The deposited material resistivity is obtained from the bar resistance (data not shown) and thickness (as measured from AFM), and equal respectively $\rho_{IBID}$=4 $\mu\Omega.m$ and $\rho_{EBID}$=0.2 $\Omega.m$. The IBID deposited material resistivity falls fairly close to Pt resistivity $\rho_{Pt}$=0.106 $\mu\Omega.m$ while the EBID material resistivity is much higher due to a lower Pt content. Such characteristics are consistent with previous reports [13,17].

We now focus on the halos visible around the deposited patterns in the SEM images of Fig. 1a (IBID) and Fig. 1b (EBID). These halos have been attributed to the surface decoration during the IBID or EBID process [18], leading to leakage pathways already at the 10 µm scale [13]. We here characterize the leakage current associated with these halos *at the sub-µm scale*. This is illustrated in Fig. 1c (IBID) and Fig. 1d (EBID) in which a series of electrodes have been fabricated, with a central gap of length 0.1 µm, 0.2 µm, 0.8 µm and 2 µm (IBID), and 0.05 µm, 0.5 µm, 1 µm and 2.3 µm (EBID). The gapped electrodes have been fabricated with the FIB operating conditions given previously. The gap resistances are plotted in Fig. 1e as a function of the gap length. IBID halos are fairly conductive (resistance lower than a few tens of k$\Omega$ for gaps below 1 µm). Such leakage values are not suitable to probe transport properties of µm or sub-µm devices (*e.g.* in carbon nanotube devices [19,20]). In contrast, the resistance of EBID halos is found between 10 M$\Omega$ and 100 G$\Omega$ for gaps with length less than 500 nm, which is compatible with transport measurement of sub-µm devices.

To support the conclusion that halos surrounding the fabricated patterns are metallic-like, we used Kelvin Force Microscopy to image the electrostatic potential distribution over a multiwalled carbon nanotube (MWCNT) device connected with IBID electrodes as shown in the schema of Fig. 2a. Fig. 2b shows the SEM image of this device. It is already visible from Fig. 2b that the halos around three electrodes (B, C and D) overlap each other, while one electrode (A) appears to be disconnected from its nearest neighbour (B). This is confirmed by the KFM image of Fig. 2c, in which the MWCNT device is imaged with the substrate and electrode A set at ground, the electrode D biased at +3 V, and the electrodes B and C unconnected. In the KFM image of Fig. 2c, dark contrasts correspond to surface potentials close to 0V, and bright contrasts to positive electrostatic potentials. The three electrodes B, C and D appear to be at the same potential in the KFM image, which is confirmed by the cross-section shown in Fig. 2d, in which the voltage drop only occurs between the electrodes with disconnected halos A and B [21]. KFM measurements thus confirm that the use of IBID is clearly limited by the presence of parasitic halos surrounding the metal deposition, and that IBID is restricted to measurements above the 10 µm scale [13]

The leakage through halos is finally further confirmed using scanning-probe measurements. IBID electrodes with a 4.5 µm gap have been investigated using conducting-AFM (see Figure 3a), in which one of the electrodes (B) is biased at +8V while the other electrode (A) and the metallized AFM tip are left at ground. The measurement consists in recording the sample topography (here, in tapping mode, see Fig. 3b) and then mapping the current which passes through the AFM tip (see Fig. 3c), when the tip is scanned in contact mode over the sample, here with a few nN contact force. Results show that the IBID halos exhibit a noticeable conduction. The local resistance through the tip is here of a few tens of $M\Omega$ at a distance ~1 µm from the biased electrode (B). This is much lower than the leakage resistance through IBID gaps in Fig. 1 at the same distance (typically a few tens of $k\Omega$), as due to the tip contact size and/or contact resistance. However, the leakage resistance however exhibits a sharp exponential-like increase of typically one decade per µm, as for the IBID gaps in Fig. 1. Thus we are confident that the two experiments correspond to the same conduction process.

We finally illustrate the possibility to electrically characterize devices contacted by EBID at the µm-scale [22]. A SEM image of an EBID contacted multiwalled carbon nanotube is shown in Fig. 4a. To circumvent the high resistance of the EBID leads, the device as then been measured using a four probe measurement scheme, in which one measures the voltage drop ΔV between the two internal leads while using the two external leads to pass a current I through the nanotube. The I(ΔV) characteristic is shown in Fig. 4b. It corresponds to a metallic behaviour with a resistance of ~500 kΩ consistent with multiwalled nanotubes probed with EBL-defined contacts [23]. This demonstrates that EBID contacts can be used to probe transport at the μm scale.

In conclusion, we have evaluated in this Letter the direct patterning of electrical contacts at the sub-μm scale using IBID and EBID metal deposition in a focused ion beam microscope. IBID was found unsuitable to probe transport properties of devices at the μm-scale due to conductive halos around patterned electrodes, as seen from transport, c-AFM and KFM. On the other hand, EBID is demonstrated as a probe of transport properties at the μm scale, however using four-point measurement schemes to account for the deposited electrode higher resistivity. This work has been funded in part by an ANR grant N° ANR-05-JCJC-0090. We thank C. Boyaval for technical assistance.

levels 4 to 6 orders of magnitude lower than for IBID electrodes, as deduced from Fig. 1, and have not been attempted so far.

**FIGURE CAPTIONS**

*Figure 1: (a) SEM image of an IBID Pt pattern bridging predefined metallic contacts fabricated by EBL. The scale bar is 2 µm. (b) Same image for an EBID Pt pattern. The scale bar is 2 µm. (c) Series of gaps defined by IBID. Gap lengths are (from top to bottom): 0.1 µm, 0.2 µm, 0.8 µm and 2 µm. (d) Series of defined by EBID. Gap lengths are (from top to bottom): 0.05 µm, 0.5 µm, 1 µm and 2.3 µm. (e) Gap resistance plotted as a function of gap length.*

*Figure 2: (a) Schema of the device, showing a multi-walled carbon nanotube, connected in four points. The two external electrodes (A and D) are used for applying a 3V bias, the two internal remaining unconnected. (b) SEM image of a multiwalled carbon nanotube contacted by four Pt electrodes deposited by IBID techniques. (c) Kelvin Force Microscopy image of the device. The scan frame corresponds to the black square in (b). Electrodes and the nanotube have been drawn as a guide for the eyes. (d) Cross-section of the surface potential corresponding to the line in (c) between the electrodes (A) and point (D).*

*Figure 3: (a) SEM image of IBID Pt patterns creating a 4.5 µm gap. The scale bar is 4 µm. (b) AFM image and (c) conductive-AFM image of the device. The scan frame corresponds to the black square in (a). (d) Cross-section of the topography and the tip-current along the blue lines respectively in fig. 3b and fig. 3c.*

*Figure 4: (a) SEM image of a multiwalled carbon nanotube (diameter 10 nm) contacted by four Pt electrodes fabricated by EBID. (b) Current I flowing through the nanotube as a function of the voltage drop ΔV measured between the two internal electrodes. The red line corresponds to a linear fit. The device resistance is ~500 kΩ.*

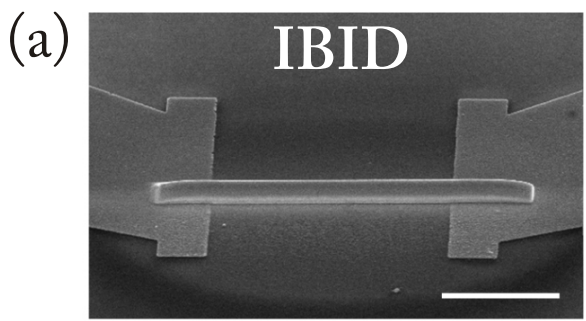
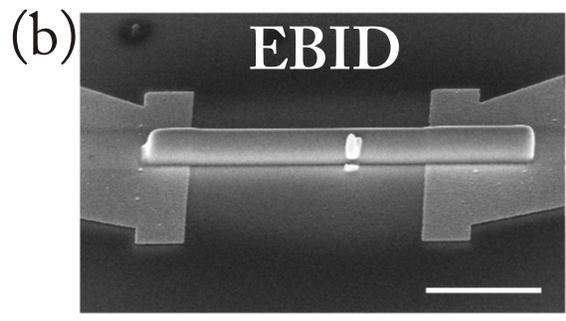
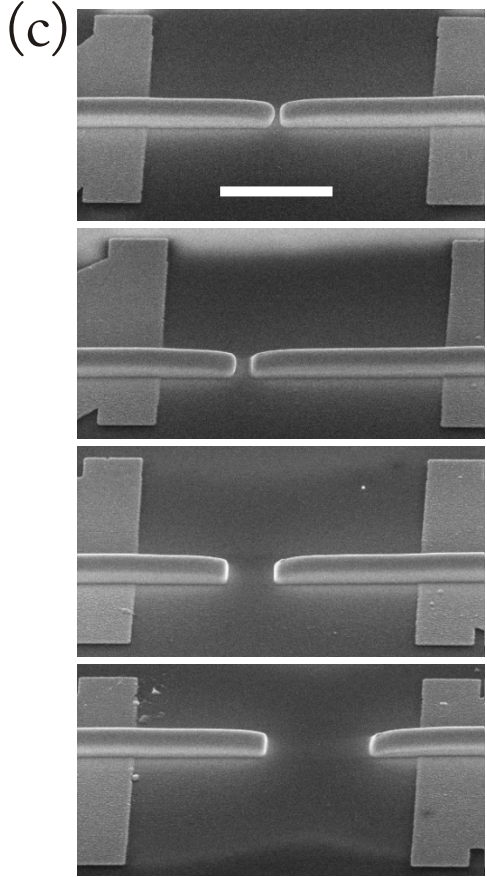
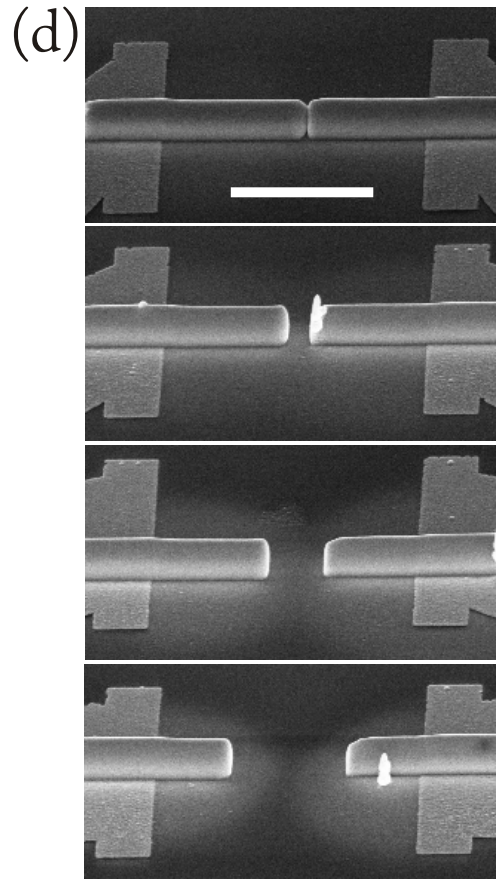
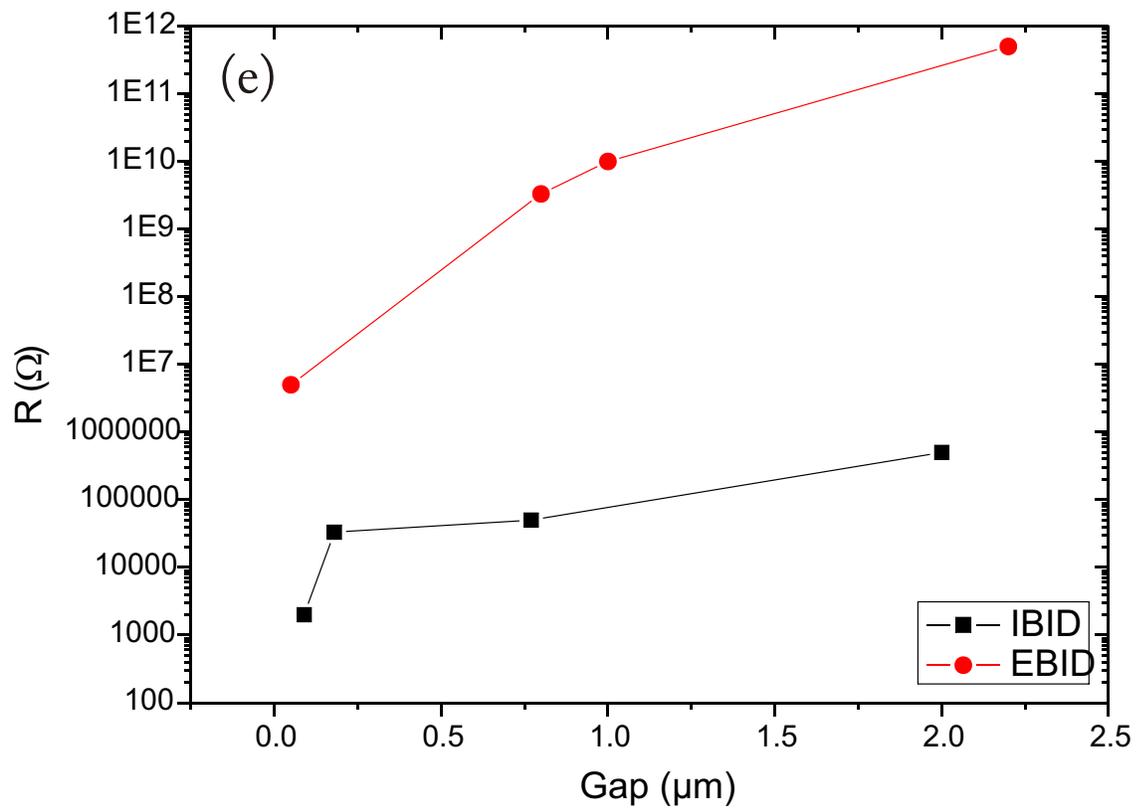

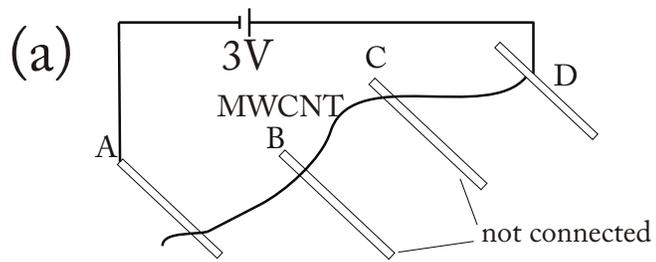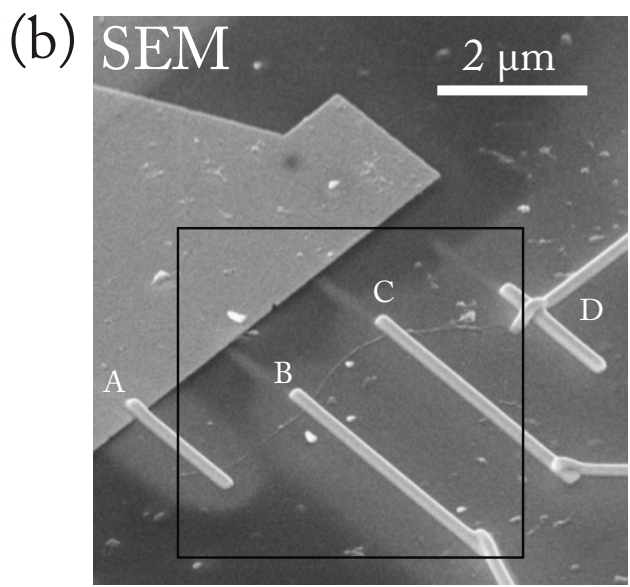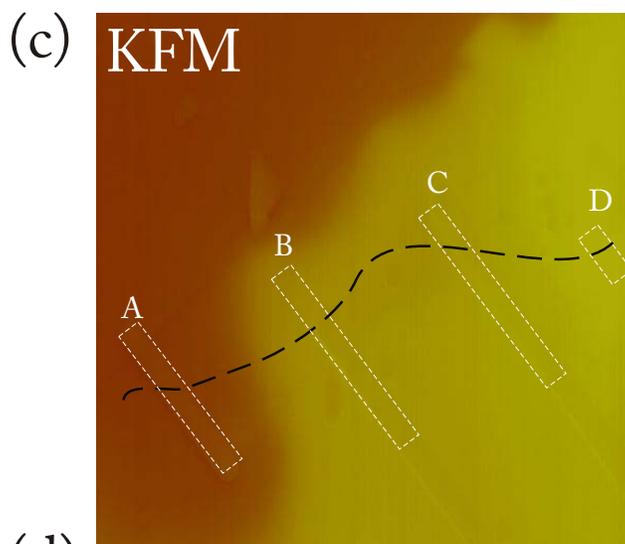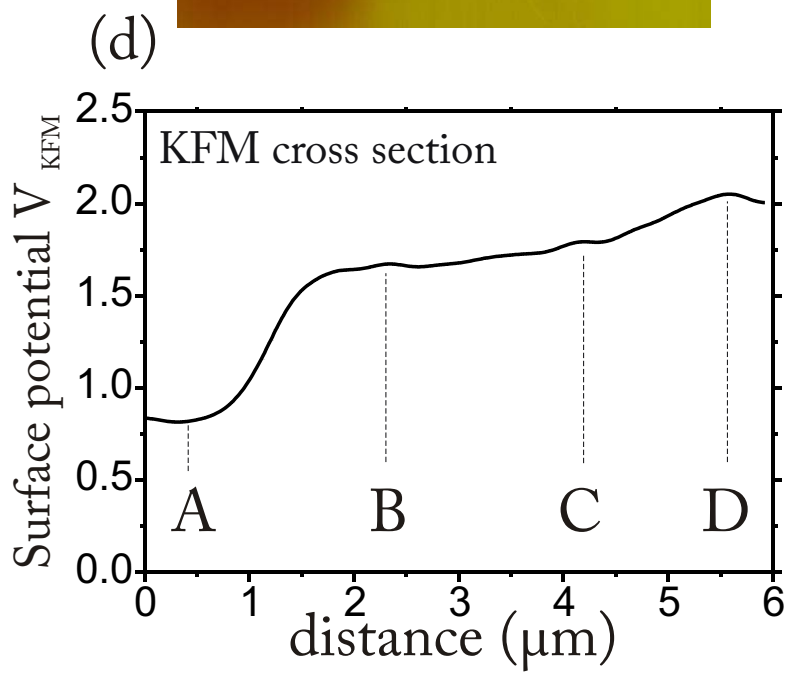

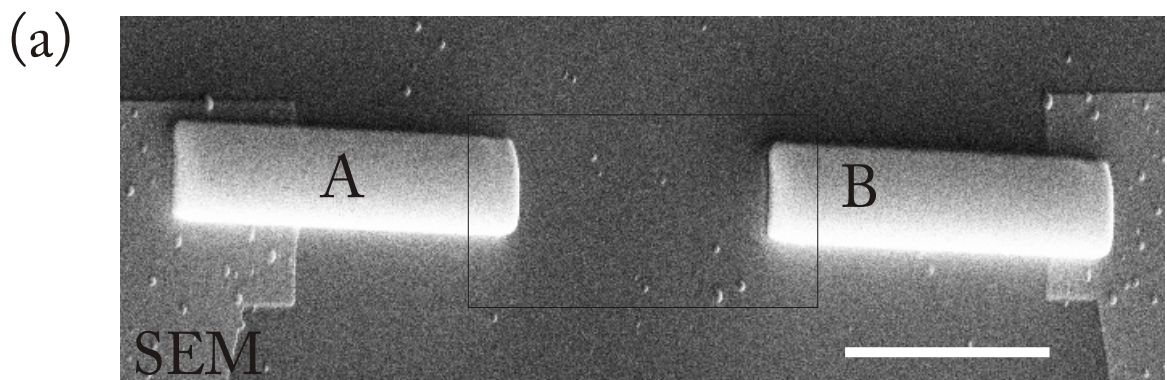
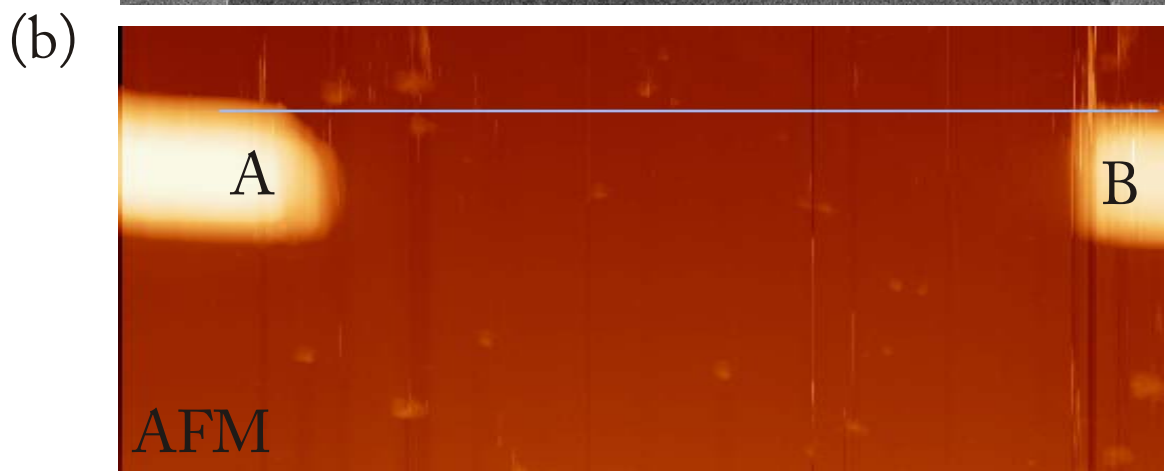
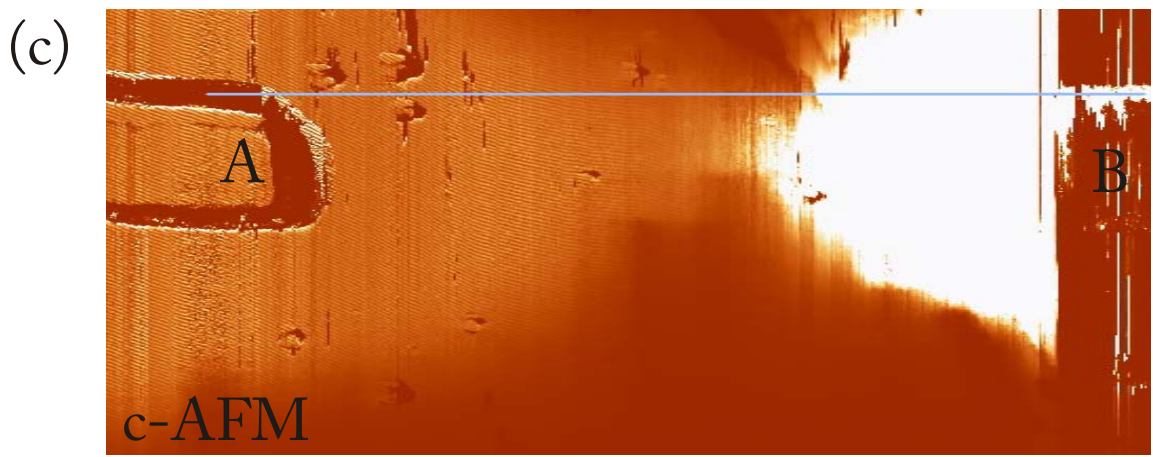
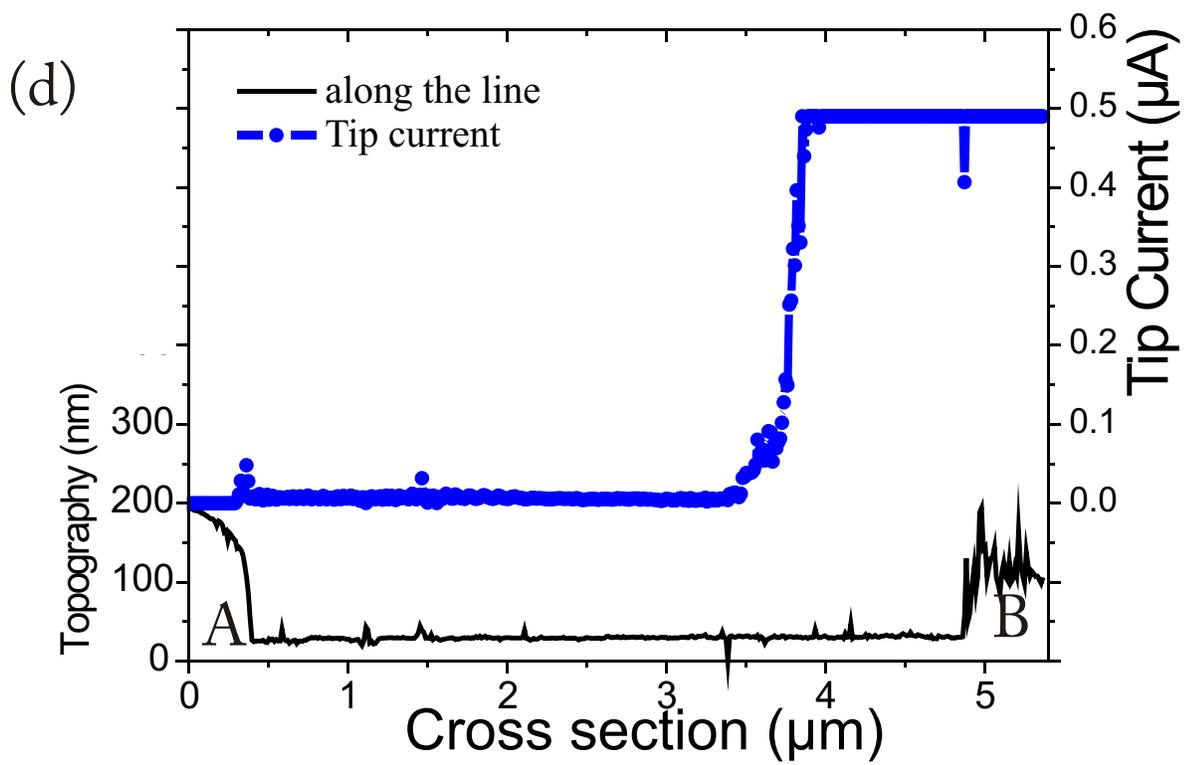

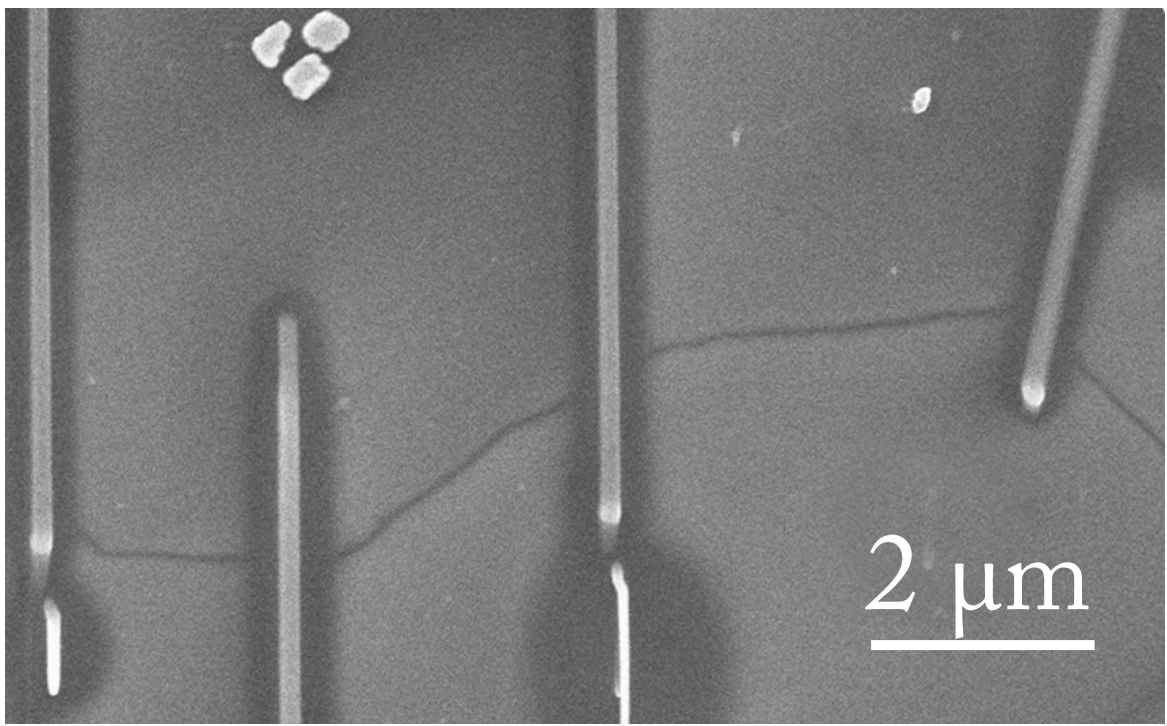

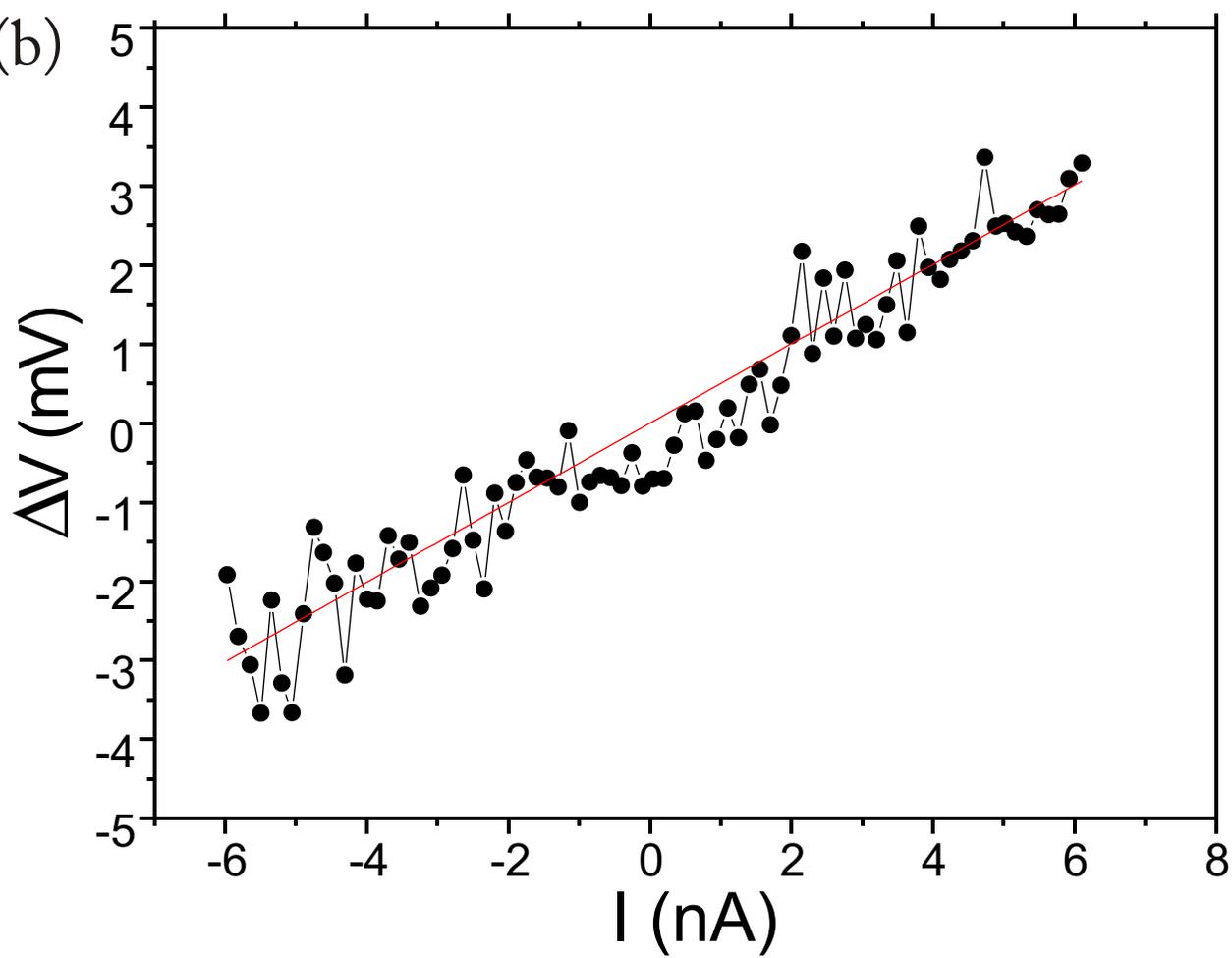